\documentclass[pdflatex,sn-mathphys,iicol]{sn-jnl}

\jyear{2021}%

\begin{document}

\title[Article Title]{Materials under high pressure: A chemical perspective}

\author[1]{\fnm{Katerina P.} \sur{Hilleke}}\email{khilleke@gmail.com}

\author[1]{\fnm{Tiange} \sur{Bi}}\email{tiangebi@buffalo.edu}

\author*[1]{\fnm{Eva} \sur{Zurek}}\email{ezurek@buffalo.edu}

\affil*[1]{\orgdiv{Department of Chemistry}, \orgname{State University of New York at Buffalo}, \orgaddress{\city{Buffalo}, \postcode{14260-3000}, \state{NY}, \country{USA}}}

\abstract{At high pressure, the typical behavior of elements dictated by the periodic table -- including oxidation numbers, stoichiometries in compounds, and reactivity, to name but a few -- is altered dramatically. As pressure is applied, the energetic ordering of atomic orbitals shifts, allowing core orbitals to become chemically active, atypical electron configurations to occur, and in some cases, non-atom-centered orbitals to form in the interstices of solid structures. Strange stoichiometries, structures, and bonding motifs result. Crystal structure prediction tools, not burdened by preconceived notions about structural chemistry learned at atmospheric pressure, have been applied to great success to explore phase diagrams at high pressure, identifying novel structures in diverse chemical systems. Several of these phases have been subsequently synthesized. Experimentally, access to high-pressure regimes has been bolstered by advances in diamond anvil cell and dynamic compression techniques. The joint efforts of experiment and theory have led to startling successes stories in the realm of high-temperature superconductivity, identifying many novel phases -- some of which have been synthesized -- whose superconducting transition approaches room temperature. }

\keywords{Pressure, crystal structure prediction, superconductivity, electronic structure}

\maketitle

\section{Introduction}\label{sec1}

Our chemical knowledge and intuition, which has accumulated from our first chemistry classes and experience living at atmospheric conditions, can be completely knocked askew when the pressure variable is taken into consideration.  The range of pressures in our universe is wide, from 10$^{-32}$~atm in the vast spaces between galaxies all the way up to 10$^{+32}$~atm in the core of a neutron star. In the core of our own planet, Earth, pressures can reach up to $\sim$3.5$\times 10^6$~atm (350~GPa), while many giant planets have core pressures in the TPa regime. Understanding how chemistry changes under pressure is key to uncovering how materials behave in these environments. Experimental techniques based on diamond anvil cells (DACs) can access static compression in the multi-megabar regime \cite{1,Dubrovinsky:2015,Jenei:2018,Dubrovinskaia:2016,Dewaele:2018}, while even higher pressures are accessible via dynamic compression using shock wave or ramp techniques. \cite{2,Hansen:2021,Fratanduono:2021} 

However, producing these high pressures environments and analyzing the resulting data is difficult.~\cite{Zurek:2014i} Light elements tend to diffuse into the diamonds of the DAC cell, weakening and breaking the apparatus, which is financially expensive and wastes research time. Unintended side reactions within the DAC can result in decompositions, unwanted product formation, and a mixture of byproducts alongside the actual synthetic target. Even following a successful synthesis, characterization is complex, since the capabilities of X-ray diffraction (XRD) are limited for light elements. The pressure limit of neutron diffraction is not high enough for many experiments, causing the indirect-probe techniques of Raman and IR spectroscopy to step into the gap. These difficulties have led to the development of a symbiotic relationship between theoreticians and experimentalists, in which theoretical calculations into promising novel compounds provide plausible synthetic targets, and aid in structural characterization and analysis of properties once experimental phases are observed. H$_3$S and LaH$_{10}$, two high-pressure phases with record-breaking superconducting critical temperatures, $T_c$, have both been investigated in such a symbiotic partnership. \cite{Strobel:2011a,Duan:2014,Li:2014,Drozdov:2015a,geballe2018synthesis,drozdov2019superconductivity,Pers105,Peng:2017a,Liu:2017-La-Y,Kim:2008-Al}

As is evident from the strange elemental ratios in the aforementioned iconic superconductoring hydrides, pressure can have a profound effect on the compositions, structures, properties, and stabilities of solid phases. \cite{Hemley:2000,Song:2013a,Grochala:2007a,Goncharov:2013b,Bhardwaj:2012a,Dubrovinsky:2013a,McMillan:2013a,Klug:2011a,Naumov:2014a,Zurek:2014i,Zurek:2016b,Hermann-lip,Zurek:2019k,Zurek:2021k} Experimentally observed compounds often could not have been predicted using the rules and bonding schemes that stem from our ambient-pressure-taught intuition. For these reasons, methods based on data mining \cite{15} and chemical intuition \cite{16}, tend to break down when predicting solid phases under pressure. Therefore, theoreticians have made use of crystal structure prediction (CSP) techniques, which aim to pinpoint the global minimum as well as interesting local minimum phases on the potential energy surface (PES) at a chosen pressure and temperature. Once the unit cells and atomic configurations -- often quite alien to those seen at ambient pressure -- are identified, theoreticians can calculate the structural and electronic properties of the resulting phases, either to compare with or to guide experimentalists. Theoretical calculations can also probe pressures beyond those which are currently accessible experimentally. 

Herein, we first give an overview of several commonly used CSP algorithms, followed by a survey of the unique behavior of matter compressed under extreme pressures. We start by outlining the changes in electronic structure that are induced by pressure, tracking how they manifest in strange stoichiometries, structures, and bonding patterns in the phases that result. Finally, we discuss the exciting superconducting phases that have emerged in high-pressure experiments.

\section{Crystal Structure Prediction}\label{sec2}

The problem tackled by CSP is one of global optimization. Within this framework, the unit cell parameters and atomic positions -- in other words, the geometric coordinates of the crystalline structures -- are the parameters that are varied towards achieving the goal of minimizing the free energy. \cite{18,19,20,21,22,23,24,25,26,27,28,29} These parameters include three unit cell vectors, three unit cell angles, and 3$N - $3 degrees of freedom in atomic positions, resulting in a total of 3$N + $3 parameters that need to be determined to identify a global minimum structure in a PES for a phase with $N$ atoms in the unit cell. The resulting problem is very difficult, for several reasons. For one, it has been shown that the number of local minima in the PES increase exponentially with the number of atoms in the unit cell. \cite{30} Secondly, the computational cost of optimizing structures to the nearest local minimum can be quite expensive, especially for the (typically) thousands of structures that need to be considered in a single run. For this reason, early CSP studies employed empirical potentials for small solid systems such as Li$_3$RuO$_4$ \cite{31} and molecules including fullerene clusters. \cite{32} Therefore, CSP is an NP-hard (nondeterministic polynomial time hard) problem, since locating the global minimum of homogeneous and heterogeneous systems both \cite{55,56} has no algorithm that scales as a polynomial in the possible degrees of freedom. 

Taking all of this into account, common meta-heuristic algorithms have been designed to search for good (low-energy, stable) structures, but they are not guaranteed to pinpoint the global minimum due to the sheer complexity of the problem. However, it can be fruitful to analyze even the local minima that are identified from such searches because they may possess promising properties, and many are synthetically accessible. Various CSP methods have been designed to sample the PES in different ways based on the situation at hand.

Some CSP methods, including random structure searches \cite{23,37}, evolutionary (EA) and genetic algorithms (GA) \cite{38,39,40,41,42,43,44,45}, and particle swarm optimization (PSO), \cite{22,46} search throughout the entire PES. If minimal to no information is known regarding the target structure(s), this type of method is preferred, because they start by sampling the entire PES to identify promising regions and only then narrow their focus. Other methods, such as metadynamics \cite{35}, simulated annealing \cite{36}, basin hopping \cite{33}, and minima hopping \cite{34}, thoroughly explore only a particular, selected region of the PES by overcoming energy barriers for a given starting structure, and therefore multiple runs with various starting phases might be necessary. Hybrid methods combining the aforementioned techniques also exist. \cite{47,48} Since interatomic potential parameters are often unreliable for squeezed crystalline structures, a common practice for high pressure research is to couple CSP with density functional theory (DFT) for local optimizations that relax the crystal structures to the nearby minima. Some methods tackle this problem by concurrently fitting DFT results produced during the search to update parameters on-the-fly. \cite{49,Tong:2018,Deringer:2018,Podryabinkin:2019,Tong:2020}

In order to acquaint the reader with various CSP techniques, we will next describe the details of several popular methods that were mainly developed for small inorganic crystals where each atom is treated independently. Although the methods can be modified to include constraints on inter-molecular interactions, \cite{50} that would necessitate taking into consideration flexibilty in conformation, nearly iso-energetic polymorphs of molecular crystals, and large unit cells, \cite{51} rendering it beyond the scope of the review.  Other CSP algorithms not included in the discussion herein are the ones for large systems at ambient conditions involving empirical information, data mining or machine learning. \cite{15,52,53,54}

\subsection{Potential Energy Surfaces}\label{subsec1}

\begin{figure}
\begin{center}
\includegraphics[width=1\columnwidth]{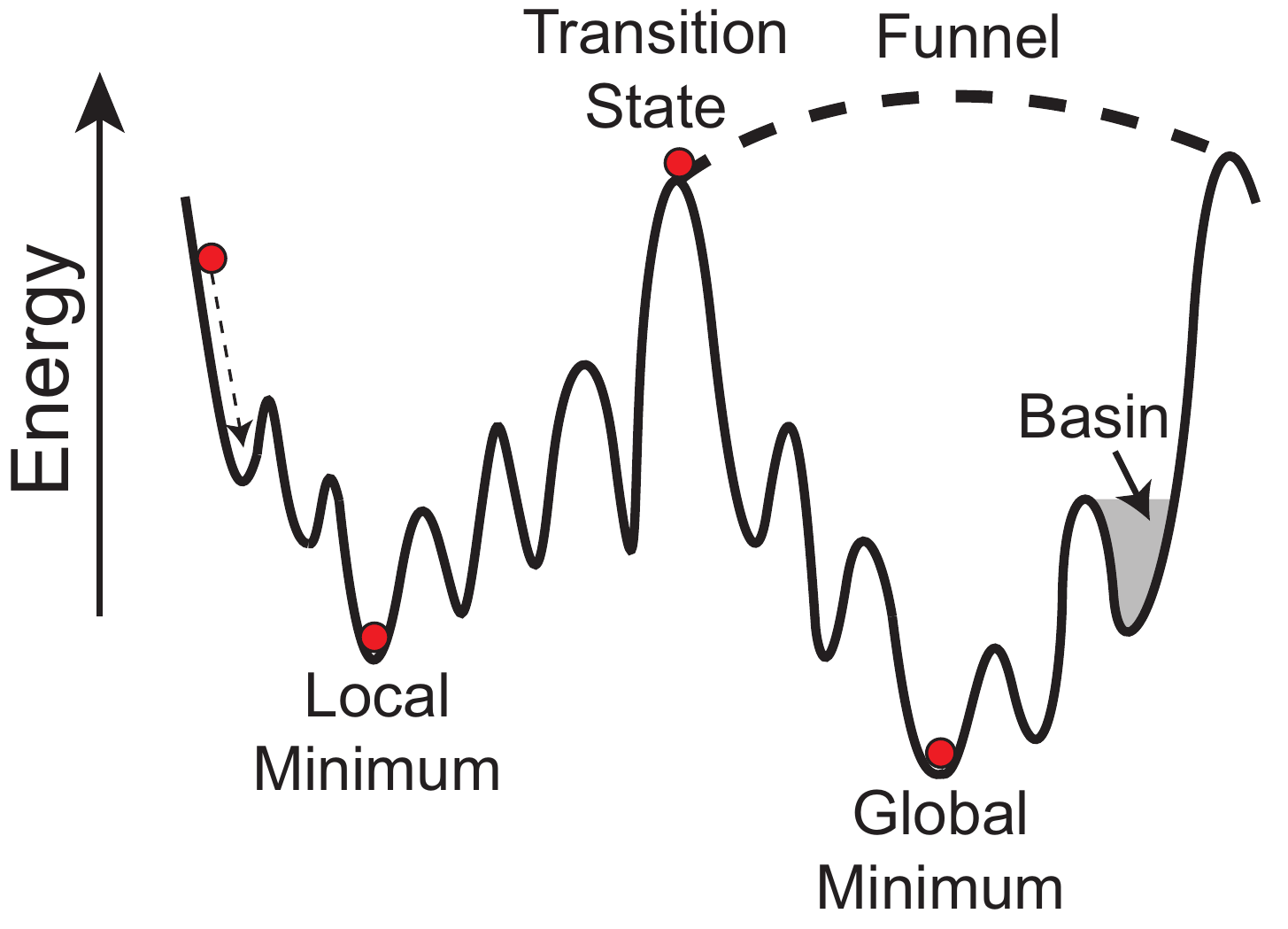}
\end{center}
\caption{Two-dimensional schematic of a potential energy surface, with red dots indicating single structures at various locations. One dot sits at the global minimum, the lowest energy point across the entire PES, while anoter sits at a local minimum in the PES. A third sits atop a peak in the PES representing a transition state between two funnels that each contain many local minima. A fourth  will, upon local optimization represented by a dotted line, go to the local minimum of the basin in which it lies. Adapted with permission from Zack, F.; Avery, P.; Wang, X.; Hilleke, K. P.; Zurek, E. \emph{J.\ Phys.\ Chem.\ C} 2021, 125, 1601. \cite{XtalOpt-2020} Copyright 2021 American Chemical Society.}
\label{fig:pes}
\end{figure}

Potential energy surfaces or landscapes are highly dimensional, describing how the energy of a system varies as a function of the 3$N+$3 degrees of freedom. PESs are decorated with numerous local minima corresponding to potential configurations of the $N$ atoms within some unit cell. The local minima manifest as ``valley"-like shapes in the PES, and they are separated by energy barriers. Each structure defined or generated by a CSP algorithm typically undergoes a process termed local optimization, where its energy is minimized to reach the nearest local minimum. The PES may contain various features, illustrated in Figure~\ref{fig:pes}: 1) basins, within which all structures will optimize to a single local minimum, 2) super basins, the domains in the potential energy surface with multiple basins located next to each other, and 3) funnels, super basins with multiple local minima that are separated by small barriers.

Locating the global minimum of large and complex systems can be very difficult. The ``no free lunch'' theorems conjecture that the success rates of all algorithms will be the same when averaged over all PESs, \cite{57} meaning that varying the method employed will not necessarily help to improve the search for a particular system. However, certain features of the PES can be leveraged to the advantage of the researcher, even for a structure with many atoms in a complex unit cell.

For example, there is no need to sample the entire PES, since the chemically desired (physically reasonable) structures are typically located only in certain areas, while the rest of the surface involves configurations with high energy structures where atomic distances are too close or too far for them to be energetically competitive. Removing these undesired regions from the searches greatly accelerates the CSP run. Another property of the PES is that low energy configurations occupy the largest amount of ``hyperspace'' in a PES, \cite{58} so that randomly generated structures tend to fall into  low energy basins rather than regions with high energy. Finally, previous studies suggested that low-energy basins are more likely to be located close to each other, with the energy barriers between them being small. \cite{59,60} Thus, a configuration falling in a random place in a funnel can be lightly perturbed, allowing it to overcome the energetic barriers and continually move to lower energy until it reaches the global minimum within the funnel. Overviews of the commonly used CSP techniques that have been used to predict the structures of high pressure materials are given below. 

\subsection{Following Imaginary Phonons}\label{subsec2}

If the target compound resembles a known structure -- a distortion of a highly symmetric parent phase as in, for example, families of perovskites, one way to carry out CSP is simply to follow imaginary phonon modes. Following local optimization of a given starting configuration and calculations of its phonon band structure, the presence of imaginary phonon modes indicate that a lower-energy phase in the PES could be accessed by displacing atoms along the eigenvectors of the soft phonon modes. The structure is modified with the displaced atoms -- typically resulting in a lower-symmetry structure -- and a subsequent local optimization is performed. The process needs to be repeated until the final structure is located at a local minimum, with no imaginary phonon modes. One special case that can occur is when the softest mode is located at the $\Gamma$ point, in other words, the center of the Brillouin zone. In this case, the perturbed structure would have the same unit cell as the original but with the atoms moved along the displacement vectors. Following these sorts of structural pathways allows us to enrich our chemical knowledge, by getting a better of understanding of the relationships between various structure types, and how symmetry breakings affect electronic properties and energies.

Extensive predictions of complex systems have been done by following imaginary phonon modes, and many of them show interesting properties. For example, distortions along soft phonon modes from a rock-salt lattice in late transition metal oxides lead to complex crystal structures \cite{64} in which the distortions are coupled either to magnetic ordering or Jahn-Teller effects. Several structures explored for their promise as high-temperature superconductors were identified by following soft phonon modes, including silane~\cite{66,65,67} and a derivative of H$_3$S intercalated wth methane~\cite{CSH7}. Distorted perovskite phases based on concerted tilting or rotations of octahedra are frequently demonstrated to arise from atom displacements along imaginary phonon modes in a high-symmetry cubic structure\cite{61,62,63,Kaczkowski:2021}. High-pressure conversions between zircon, scheelite, and fergusonite structures in MXO$_4$ phases including CaMoO$_4$\cite{Panchal:2017} and YVO$_4$\cite{Mittal:2019} can also be understood through this lens.

Although this method draws a clear route from inital structure to final answer, there are limitations: the original structure should be close to the target in the PES, in other words, they should have similar configurations. Reorienting octahedra in a perovskite lattice requires little in terms of atomic rearrangement; other high-pressure structures may be far more alien. Another disadvantage is that the actual practice can be computationally expensive even within DFT, since phonon calculations must be performed across the entire Brillouin zone. 

The method of following soft phonon modes can be combined with several of the CSP techniques described below, however, by using their output to define the initial starting guess. Because CSP techniques are typically limited to small unit cell sizes, the most stable phases they find may show soft modes away from the Zone center, which serve as a guide to a related, dynamically stable compound.

\subsection{Simulated Annealing}\label{subsec3}

\begin{figure*}
\begin{center}
\includegraphics[width=2\columnwidth]{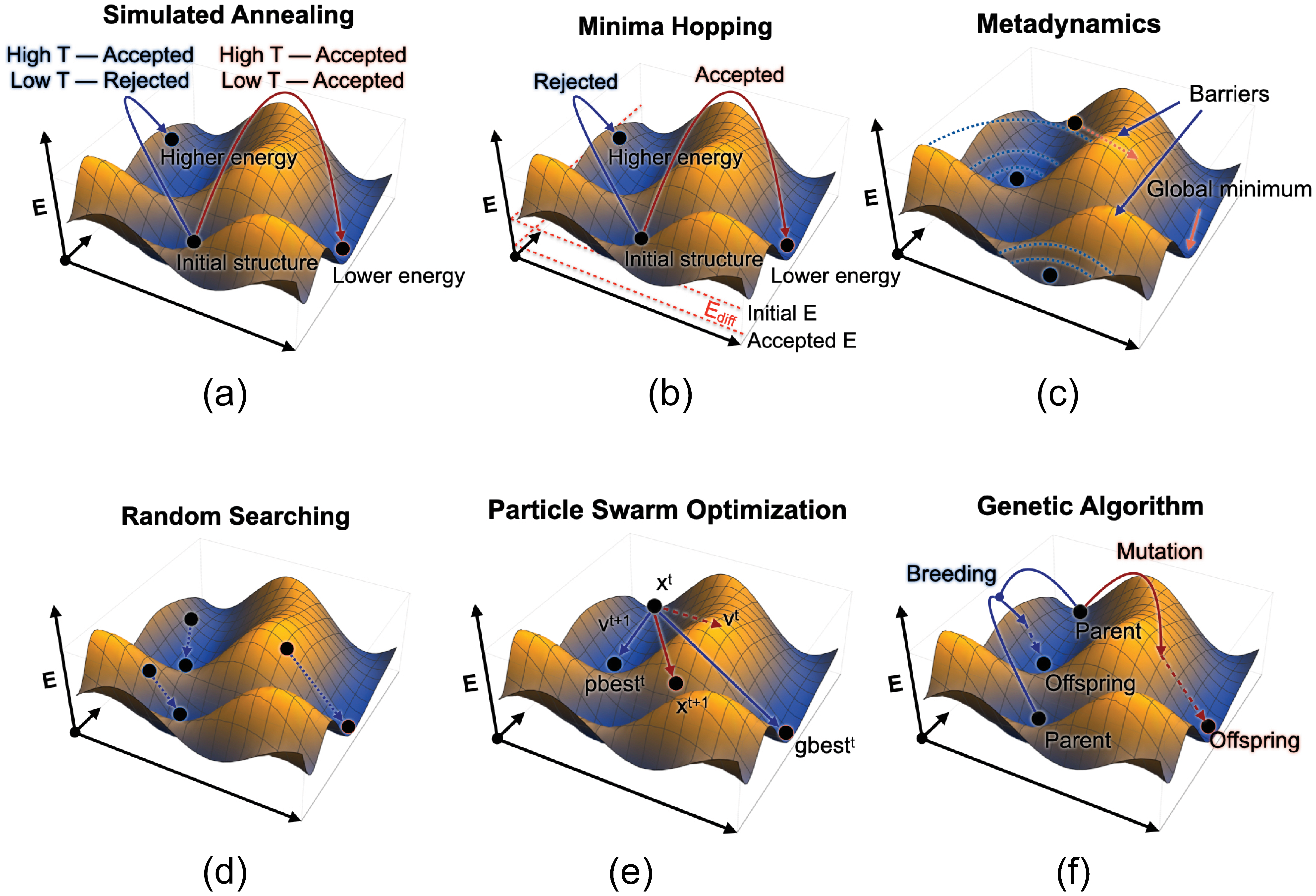}
\end{center}
\caption{Schematic illustration of the ways in which CSP techniques traverse a three-dimensional PES, including: (a) simulated annealing, (b) minima hopping, (c) metadynamics, (d) random searches, (e) particle swarm optimization, and (f) genetic algorithms. Adapted with permission from Zack, F.; Avery, P.; Wang, X.; Hilleke, K. P.; Zurek, E. \emph{J.\ Phys.\ Chem.\ C} 2021, 125, 1601. \cite{XtalOpt-2020} Copyright 2021 American Chemical Society.}
\label{fig:pes3d}
\end{figure*}

Another technique that probes local regions of the PES is simulated annealing, which was inspired by the heating and quenching process of metallurgy that recrystallizes materials \cite{36}. To start, a random configuration of atoms is generated and its energy assessed. From this starting point, a new structure is produced by some combination of perturbation, random atomic displacement, and atom permutation of the initial structure. The probability $P$ of the new structure being accepted is based on the Metropolis Monte Carlo algorithm, where $P = -\Delta E/k_BT$ ($\Delta E$ being the energy difference between the initial and new structures and $k_B$ is the Boltzmann constant). The variable $T$ is the ``simulation temperature", distinct from physical temperature, that is employed in the mathematical expression controlling the acceptance rate of new structures. After a period of ``heating", where values of $T$ are large, a random number $\epsilon$ (0 $< \epsilon <$ 1) is used as a cutoff value so that new structures will be accepted only if $\epsilon$ is smaller than $P$. As the process iterates the $T$ variable is continually decreased -- analogous to the slow cooling of a solid material from the melt to form crystals -- to tighten the criteria of structure acceptance, as illustrated schematically in Figure~\ref{fig:pes3d}a. Once $T$ is reduced to 0, the process will only accept structures with lower energies than the parent phase, locating the neighboring local minima.

The success of this method relies significantly on choosing a good starting structure. In addition, it is assumed that the high-temperature PES has a similar landscape to the low-temperature PES, so that the local minima before and after cooling are the same, but this does not always hold true. More complicated search schemes, which can involve multiple heating/cooling cycles and random mutations, have been proposed as techniques that explore broader regions of the PES. However, random mutations in simulated annealing have to be carefully handled for two reasons: 1) mutations that are too large can cause a run to sample a large portion of the PES, effectively the same as a random structure search, and 2) too-small mutations can fail to effectively perturb the system. Molecular dynamics can also be used to perturb the structures.

The application of the simulated annealing method has been fruitful, coupled in many cases to Hartree-Fock and DFT methods. \cite{20,77} Many inorganic crystal structures, including in the Pb-S systems, \cite{78} Ca-C systems, \cite{79} metal-N systems, \cite{82} B-N systems, \cite{80} and Ge-F systems, \cite{81} have been identified using this method.

\subsection{Basin and Minima Hopping}\label{subsec4}

Similar to simulated annealing, the basin and minima hopping methods rely on a judicious choice of starting structure. Basin hopping, developed mainly for finite cluster systems \cite{33,83,84,85,86}, relies on the Monte Carlo algorithm to evaluate the fitness of proposed structures to be accepted or rejected based on a carefully selected ``temperature" cutoff. In this method, all structures in a basin undergo relaxation to the same local minimum, so that the PES is transformed into a series of connected step functions that are each related to a local minimum. No penalty is imposed for sampling a region of the PES that has already been searched. 

Although the minima hopping method is very similar in design to basin hopping, it uses molecular dynamics rather than Monte Carlo to traverse the PES. As such, a temperature variable is not required. Instead, after relaxing a structure into a nearby local minimum, the cutoff energy difference between the original and optimized structures, labeled $E_\text{diff}$ in Figure~\ref{fig:pes3d}b, is set so that half of the new structures are accepted. Using soft phonon modes to define inital trajectories for the molecular dynamics step can increase the efficiency of the method,  in-line with the Bell-Evans-Polanyi principle regarding relative energies of local minima and the energetic barriers surrounding them. \cite{Roy:2008,Sicher:2011} The algorithm can also selectively favor exploring new regions of the PES by artificially raising the kinetic energy employed in the molecular dynamics  step. This mechanism is invoked when already explored regions of the PES are revisited.

High pressure phases investigated by minima hopping include various polymorphs of elemental carbon \cite{88,89}, as well as superconducting hydrides of phosphorus, \cite{90} sulfur, selenium, \cite{91}, silicon \cite{87} and ternary S$_x$Se$_{1-x}$H$_3$ hydrides \cite{Amsler:2019}, as well as to identify new high-pressure structures of SrTiO$_3$\cite{Khajehpasha:2021}. Several binary systems that are known to be immiscible at ambient pressures have been investigated with the minima hopping method, including the  intermetallic Fe-Bi \cite{Amsler:2017,Amsler:2018}, Cu-Bi \cite{Amsler:2017b,Clarke:2017,Amsler:2018}, and Ni-Bi \cite{Powderly:2017,Amsler:2018} systems.  Using a linear approximation to the enthalpy to predict the stabilities of compounds found in 1~atm databases, the minima hopping method identified stable phases in the immiscible-at-ambient-pressure systems As-Pb, Al-Si, Sn-Bi, Fe-In, Hg-In, Hg-Sn, Re-Sn, Re-Br, and Re-Ga. \cite{Amsler:2018}

\subsection{Metadynamics}\label{subsec5}

The metadynamics method is based on molecular dynamics simulations. Exploration of the full PES is encouraged by filling already-visited areas with Gaussian functions, represented by the dotted lines in Figure~\ref{fig:pes3d}c, to overcome barriers between known and unknown regions.~\cite{35} This so-called ``basin-flooding'' is typically performed in a space defined by a particular collection of variables, fewer than the 3$N+$3 needed to represent the full crystalline lattice. Metadynamics has been frequently applied to studying phase transitions \cite{94,95}, structural changes in solutions, and chemical reactions, \cite{35}, but has also been employed to predict several high pressure structures, including phases of Ge \cite{92}, Ca, \cite{71}, N, \cite{Adeleke:2017} and carbon dioxide. \cite{93} One shortcoming of the metadynamics method is that it may flood a transition basin separating multiple minima, thereby making it impossible to find the global minimum. Thus, unless a good starting structure is known, it is important to perform multiple searches to ensure adequate exploration of the PES.

\subsection{Random (Sensible) Structure Searches}\label{subsec6}

The previously discussed CSP algorithms -- following soft phonons, simulated annealing, basin hopping, and metadynamics -- all excel in the fine exploration of particular regions of the PES and to some extent rely on a reasonable structural guess as a starting point. The following three metaheuristics, starting with random searches and proceeding to particle swarm optimization methods and genetic/evolutionary algorithms specialize instead in surveying a full PES to eventually narrow in on a final region. 

Of these, the random search method is rooted in the most straightforward idea: to generate crystal structures with randomized unit cell parameters and atomic coordinates, and then to relax these structures to the closest local minima, as illustrated in Figure~\ref{fig:pes3d}d. This ensures wide coverage of the full PES, not biased to any particular region. However, since randomly generated structures might contain unrealistic interatomic distances, falling into unhelpfully high energy regions of the PES, constraints can be applied to avoid atoms occupying positions that are too near or too far from one another. If experimental observations or chemical intuition confer knowledge regarding lattice parameters, types, or space groups, such constraints can also be applied. For structures containing molecules, quasi-molecules, or clusters, connectivity constraints can further assist the procedure. Therefore, random structure search methods can also be described as ``sensible structure searches". \cite{23}

The simplest random structure searches generate completely independent structures, and therefore do not learn from their history. One algorithm that can, in principle, learn during the course of its trajectory is the  \textit{ab initio} random structure searching method (AIRSS). \cite{23,37} AIRSS allows users to apply various constraints as outlined above. Users may also invoke a ``shaking" feature, which randomly mutates previously identified stable structures through atomic displacement and unit cell deformation, thereby enabling the search to overcome barriers between phases.

Random searches have been successfully used to predict many structures, including elemental phases of hydrogen \cite{69,70}, lithium \cite{71,72}, nitrogen \cite{Turnbull:2018,Sontising:2020}, and germanium \cite{Kelsall:2021}. Additionally, high pressure phases of titanium dioxide \cite{73}, silane \cite{Pickard:2006}, xenon oxides \cite{Dewaele:2016}, MoBi$_2$ \cite{Altman:2021}, as well as several high-temperature superconducting hydrides \cite{Li:2019} have been explored through random searching methods. 

\subsection{Particle Swarm Optimization}\label{subsec7}

Inspired by the behavior of swarms of animal groups (i.e.\ flocks of birds, schools of fish), \cite{96} the particle swarm optimization algorithm, \cite{46} implemented within the Crystal Structure AnaLYsis by Particle Swarm Optimization (CALYPSO) code \cite{22}, incorporates the positions and PES trajectories of the collected group of putative structures to determine the path a particular structure will take to traverse the PES. 

The first applications of the particle swarm method focussed on predicting cluster structures \cite{97}. By now, the CALYPSO code is one of the most popular methods for high pressure CSP, and has been employed to successfully predict many novel materials, including multiple superconductors with high critical temperatures \cite{98,99,100,101,Liu:2017-La-Y,Li2MgH16,Sun:2020,Chen:2021}, structures of elemental Li \cite{148}, superhard B-N binary phases \cite{Yang:2014,Zhang:2021}, Ce-F compounds, \cite{102} and hydrides of Cl. \cite{103}

In the PSO algorithm, a set of random structures are created with constraints on interatomic distances, volumes, and symmetry.  These undergo local relaxation to the nearest local minima. Each individual's path through the PES is defined both by its previous route and the traveling history of the other individuals, expressed as a position plus a velocity. The velocity, $\nu(t + 1)$, is given as

\begin{multline}
  \nu(t + 1) = \omega\nu(t) + c_1r_1(lbest(t) - x(t)) \\ 
+ c_2r_2(gbest(t) - x(t)) \label{eq1}  
\end{multline}
where $\omega$, whose range is between 0.4 and 0.9, is an inertia term that is varied on-the-fly ; $c_1$ and $c_2$ are parameters comparing the current position of the individual to the global minimum; $r_1$ and $r_2$ are random numbers falling between 0 and 1; $x(t)$ is the current position of the individual; $lbest$ is the current local minimum that the individual is relaxed into; and $gbest$ is the best position, or current global minimum, of the swarm. An example of this trajectory calculation is shown in Figure~\ref{fig:pes3d}e.

The PSO approach has the dual benefit of continually learning from the search history, while at the same time exploring the most promising regions of the PES. To ensure the search does not become too narrow, new random structures are continuously injected, thereby broadly sampling the rest of the PES.

\subsection{Evolutionary Algorithms}\label{subsec8}

In the past two decades, a plethora of CSP codes have been developed based on evolutionary (or genetic) algorithms (EAs/GAs), including \textsc{XtalOpt}, \cite{38,XtalOpt-2020,XtalOpt-Original,XtalOpt-r10,XtalOpt-r11,XtalOpt-r12,104,105} GASP, \cite{Tipton:2013} USPEX, \cite{40,106,107} MAISE, \cite{41} EVO, \cite{42}, and other codes developed by Trimarchi et al., \cite{43,108}, Abraham et al., \cite{44} Fadda et al., \cite{45}, Woodley and Catlow \cite{Woodley:2009}, Hammer et al. \cite{Hammer:2018}, and the ``adaptive-Ga" of Wentzcovitch et al. \cite{49}. These codes are coupled with periodic first-principle simulation packages or interatomic potentials for local optimizations. The evolutionary algorithm draws inspiration from the concepts of evolutionary biology, wherein selection, mutation, and reproduction operations cooperate to produce fit organisms -- in the case of CSP, global minima on a PES. The terms EA and GA are often used interchangably, even though technically the two differ in the way that offspring are created. In EAs the operators act on the structures in real space, whereas in GAs the structural variables are first mapped onto a string, akin to a chromosome, and the heredity operations are performed on these. 

The first step in an EA search is to generate a set of random structures that can be optionally constrained by a series of parameters including interatomic distances, unit cell volumes and lattice parameters, space groups, and molecular units. Optionally a search can be seeded by candidate structures, if they are known. Following local optimization into the nearest local minimum, the fitness of each structure is evaluated based on a thermodynamic variable such as enthalpy or energy, which can in certain cases be weighted alongside another variable, such as Vickers hardness \cite{XtalOpt-r12}. The fittest structures serve as parents for the following generation, passing down whichever structural features promote stability. 

Offspring can be constructed via ``breeding", or crossover, operations in which the unit cells of two parent structures are combined, or ``mutation" operations in which a single parent unit cell is perturbed in some way, as illustrated in Figure~\ref{fig:pes3d}f. The crossover operation, originally implemented for the prediction of clusters \cite{32}, involves rotating parent structures into a random orientation, then cutting both and splicing them back into a single child. Mutation operations range from permutation and exchange operators, which switch atomic positions of different elements, to ripple operators which displace atoms according to a periodic function, to strain operators which alter the unit cell parameters entirely. They can also be combined with one another - for example, the ``stripple" operation implemented in \textsc{XtalOpt} \cite{105} involves both strain and ripple processes.

The success of EA methods is demonstrated in the large number of high-pressure crystal structures they have identified. These include hydrogen-rich phases with ``non-classical" stoichiometries \cite{109,110,111,112,113,114,115,116,117,118,119,120,121}, elemental phases of boron \cite{Oganov:2009,Hilleke:2021} and sodium \cite{142}, binary systems ranging from Xe-O \cite{122} to Sc-N \cite{Aslam:2018} to Li-B \cite{123,126} and Be-B, \cite{124,125} (and the Li-Be-B ternary system \cite{127}) alongside other ternary systems including Li-F-H \cite{LiFHn} and La-B-H \cite{DiCataldo:2021}.

\section{Chemistry Under Pressure}\label{sec4}

The application of external pressure has drastic consequences for materials behavior. To a first approximation, an increase in pressure corresponds with a decrease in the volume of condensed matter, driving up the importance of the $PV$ contribution to system enthalpy in the expression $H = U + PV$. In such a framework, structures that minimize the total volume, with high coordination numbers and simple, highly symmetric atomic arrangements, should be the natural endpoint of matter under extreme compression. These trends have been encapsulated by  various ``rules'' for materials at high pressure proposed by Prewitt and Downs \cite{128}, Grochala et al.\ \cite{Grochala:2007a} and Zhang et al. \cite{Zhang:2017} Since our chemical intuition, trained at 1~atm, is hard-pressed to generate meaningful insights at high pressures, these rules have been developed from swathes of experimental data. Nevertheless, strange chemistry is afoot, as supposedly simple metals such as Na and Li adopt open framework structures \cite{142,147} and bizarre stoichiometries and structural motifs crop up. 

In addition to mechanical consequences, applied pressure influences the electronic structure of compressed matter, altering the relative energies of atomic or molecular orbitals. As a result, unfamiliar oxidation states, reactive core electrons, and non-atom centered orbitals emerge in compressed matter, manifesting in strange stoichiometries, structures, and bonding arrangements. Elements that are immiscible or inert at ambient pressure can form compounds. The resulting properties are diverse, ranging from superconductors to topological materials. 

In the following sections, we provide an overview of the many fascinating results from the high pressure research community that demonstrate the fabulous chemical and physical properties emerging under pressure. The sections are organized according to different phenomena that arise at high pressure -- but as many stem from the common root of reordered orbital energies, they are strongly coupled. Novel electronic structures result in strange stoichiometries, which also involve bizarre bonding schemes. As such, some examples will appear in multiple places, emphasizing different facets of their novelty.

\subsection{Electronic Structures}\label{subsec9}

Pressure can have a profound impact on the electronic structure of matter. Almost a century ago, Bernal proposed that all matter will become metallic upon sufficient compression -- including hydrogen. Around ten years later, Wigner and Huntington further developed this idea, proffering 25~GPa as an estimate of the pressure above which hydrogen would transform to a monoatomic metallic solid phase. \cite{130} At ambient pressure, hydrogen exists as a diatomic molecule with a large energy gap between its highest occupied and lowest unoccupied molecular orbitals. Under applied pressure, hydrogen molecules are forced closer together, and bands are formed via orbital overlap. The increased orbital overlap induced by applied pressure raises the energy of the valence and lowers the energy of the conduction bands, eventually closing the gap between them. Concurrently, the inter- and intra-molecular H-H distances equilibrate to produce an equally spaced chain of hydrogen atoms -- a state for which the top of the valence band and the bottom of the conduction band are degenerate in energy. 

\begin{figure}
\begin{center}
\includegraphics[width=1\columnwidth]{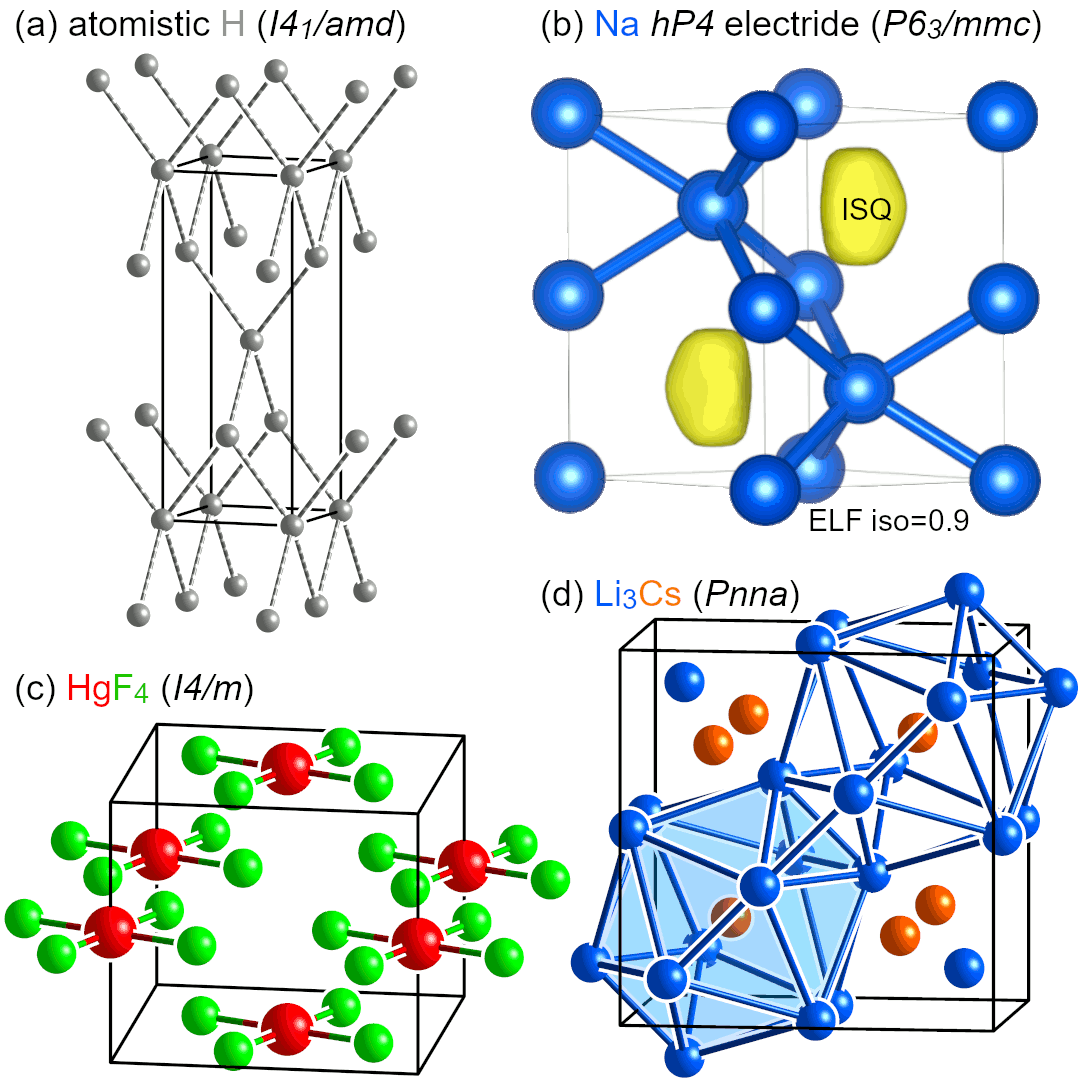}
\end{center}
\caption{High pressure alters the electronic structure of atoms in materials. Some of the fascinating compounds that have been predicted  include: (a) $I4_1/amd$ phase of metallic hydrogen, predicted ca.\ 500~GPa with monatomic H (lines do not denote bonds, but serve as a guide to the eye), \cite{70} (b) electride $hP4$ structure of Na, with ISQs present in void spaces, \cite{142} (c) HgF$_4$ where Hg assumes a +4 oxidation state, \cite{155} and (d) Li$_3$Cs, where anionic Cs takes on a charge that is less than $- 1$. \cite{156}}
\label{fig:hpreview_fig2.png}
\end{figure}

Experimental forays into metallizing hydrogen rapidly ascertained that the original suggested applied pressure of 25~GPa was a gross underestimate \cite{131,132,133,134,135}.  At the same time theoretical studies sought to ascertain the high-pressure structural transitions of elemental hydrogen \cite{Naumov:2014a,70,136,69,138,139}, uncovering complex crystal chemistry. Structures ranging from those that are layered, to ordered arrays of H$_2$ molecules to monatomic arrangements (Figure~\ref{fig:hpreview_fig2.png}a) have been identified by computations. The search for metallic hydrogen has become something of a ``holy grail" for the high-pressure research community, driving both theoretical and experimental advances.

The metallization of hydrogen is one example of how pressure affects electronic structure, and other remarkable consequences have been predicted or observed.  For instance, contrary to the expectation that under pressure all materials should assume simple close-packed structures and become metallic, several elements go rogue. Around 200~GPa, sodium transforms into an insulating and transparent phase with a band gap of 1.3~eV. \cite{142,144}. Theoretical predictions, later corroborated by experimental work, found a loss of metallicity in Li above 70~GPa \cite{145,146}. What, then, has happened here? 

As pressure increases, the electrons associated with an atom or ion occupy a smaller and smaller space, therefore their quantized energy levels increase in energy. An electron associated with a void or interstitial region in a crystalline lattice would experience the same rise in energy -- except that such an electron, unlike a valence electron of an atom, would not have to maintaining orthogonality to core electrons, thereby having less space in which it could move. The decreased space results in a higher kinetic energy. At some point, it would be energetically preferable for an electron to occupy the interstitial spaces rather than the valence atom-centered orbitals. This produces an \emph{interstitial quasiatom} (ISQ), which possesses quantized energy levels for electrons \cite{Rousseau:2008,149,Miao:2015}, and just like atom-centered electrons the ISQ can participate in bonding and chemistry \cite{Miao:2017,Miao:2015}. The ISQ-containing materials are electrides, which are well-explored under ambient conditions. \cite{Dye:2009} In the case of Na, under sufficient pressure its 2$p$ core orbitals start to overlap, squeezing its valence 3$s$ electrons into the interstitial sites (Figure~\ref{fig:hpreview_fig2.png}b) and rendering the phase insulating, \cite{142,144} while in the case of Li it is the 2$s$ electrons that are localized into interstitial sites, with an array of structures predicted. \cite{71,72,147,148,145}

A systematic investigation into the propensity towards ISQ formation used a helium ``compression chamber'' comprised of a supercell fcc lattice of He atoms whose unit cell parameters could be varied to simulate pressure. In this model the central atom was either left empty (to model an ISQ) or replaced with another element.\cite{149} The orbital energies of the ISQ model could then be compared with those of the lone, compressed atom. Li and Na were found to favor the formation of an ISQ at the lowest pressure of any element, followed by Al and then Mg at much higher pressures. This study found that ISQ formation is favored for elements with lower ionization potentials (reflecting the potential facile loss of  an electron to an interstitial region), with relatively incompressible core electrons. Moreover, because the energies of vacant $d$ orbitals do not increase as rapidly as occupied $s$ orbitals, ISQ formation was stymied via $s\rightarrow d$ electronic transfer in elements such as Cs. \cite{149}

This different rates of energy increase with pressure for atomic orbitals has consequences beyond favoring or disfavoring electride formation.  Under pressure, $s \rightarrow p$, $s \rightarrow d$, $s \rightarrow f$, and $d \rightarrow f$ electronic transitions drive strange behavior across the periodic table.\cite{Rahm:2019} For example, cesium becomes a d$^1$ metal by 15~GPa. \cite{150,151} With continuing pressure increases, the energetic reordering of Cs atomic orbitals causes a sequence of phase transitions until reaching Cs IV, a non-close-packed tetragonal structure, stable from 4.3 to 12~GPa. \cite{152}. Similar to what we saw for Li and Na, Cs IV is categorized as an electride, with the maxima in the valence charge density found in the interstitial sites. Above 70~GPa, a close-packed structure resurfaces with hybridized 5$p$ (semi-core) and 6$d$ orbitals near the Fermi level. \cite{154} 

The pressure-induced orbital reordering can also result in unusual oxidation states. Typically at ambient pressure, the oxidation states assumed by each element are dictated by their position on the periodic table relative to the noble gases -- gaining or losing only so many electrons as needed to obtain a closed-shell configuration (or a half-filled d shell). Under pressure, though, the oxidation states of mercury are predicted to diverge from the typical $+$2 to $+$3 and $+$4 in the mercury fluoride phases of HgF$_3$ and HgF$_4$ (Figure~\ref{fig:hpreview_fig2.png}c) at 40 and 75~GPa, respectively. \cite{155}  In HgF$_4$, two Hg 5$d$ electrons are used in forming bonds with four neighboring F atoms along with the usual 6$s$ valence electrons. Other predicted examples are the Li$_{1-5}$Cs phases, in which the Cs atoms assume an oxidation state beyond $-$1 due to electron transfer from Li 2$s$ to the Cs 5$d$ orbitals. \cite{156} The most negative charge on Cs is seen in Li$_3$Cs (Figure~\ref{fig:hpreview_fig2.png}d). Moreover, a series of CsF$_n$  compounds wherein the Cs 5$p$ orbitals bond with the F 2$p$ orbitals, allowing Cs to access charges larger than +1 have been studied. \cite{102} The atypical electronic configurations available under pressure result in curious stoichiometries and structural motifs in these compressed phases.

\subsection{Novel Stoichiometries}\label{subsec10}

At atmospheric pressure, one can typically deduce the stoichiometry of a binary compound using typical oxidation states to deduce their relative ratios, e.g.\ Na$^+$ and Cl$^-$ will form NaCl. However, under pressure the stoichiometry of compounds can become very different from what is observed at ambient conditions, and some elements that are immiscible or nearly so at 1~atm might undergo compound formation. 

For example, pressure opens to binary combinations of sodium and chlorine a range of stoichiometries beyond the simple 1:1: Na$_3$Cl, Na$_2$Cl, Na$_3$Cl$_2$, NaCl$_3$, and NaCl$_7$ have all been predicted to form under pressure, and Na$_3$Cl and NaCl$_3$ have been synthesized. \cite{171} As described in the previous section, cesium is a component of multiple compounds with atypical stoichiometries. At atmospheric pressure, CsF forms in the rock salt structure, with formal charges of $+$1 for Cs and $-$1 for F. Under pressure, computations have shown that the Cs 5$d$ orbitals become higher in energy than the F 2$p$, allowing Cs to attain oxidation states up to $+$3 in CsF$_3$ and $+$5 in CsF$_5$.\cite{102,Miao:2017b} CsF$_2$ and CsF$_4$ stoichiometries with Cs in $+$2 and $+$4 oxidation states have also been predicted. \cite{164} With Li, Cs has been computed to form Li$_{1-5}$Cs phases, with Cs taking on the anionic role -- in line with predicted changes in the relative electronegativities of the alkali metals under pressure. \cite{156,Rahm:2019}

\begin{figure}
\begin{center}
\includegraphics[width=1\columnwidth]{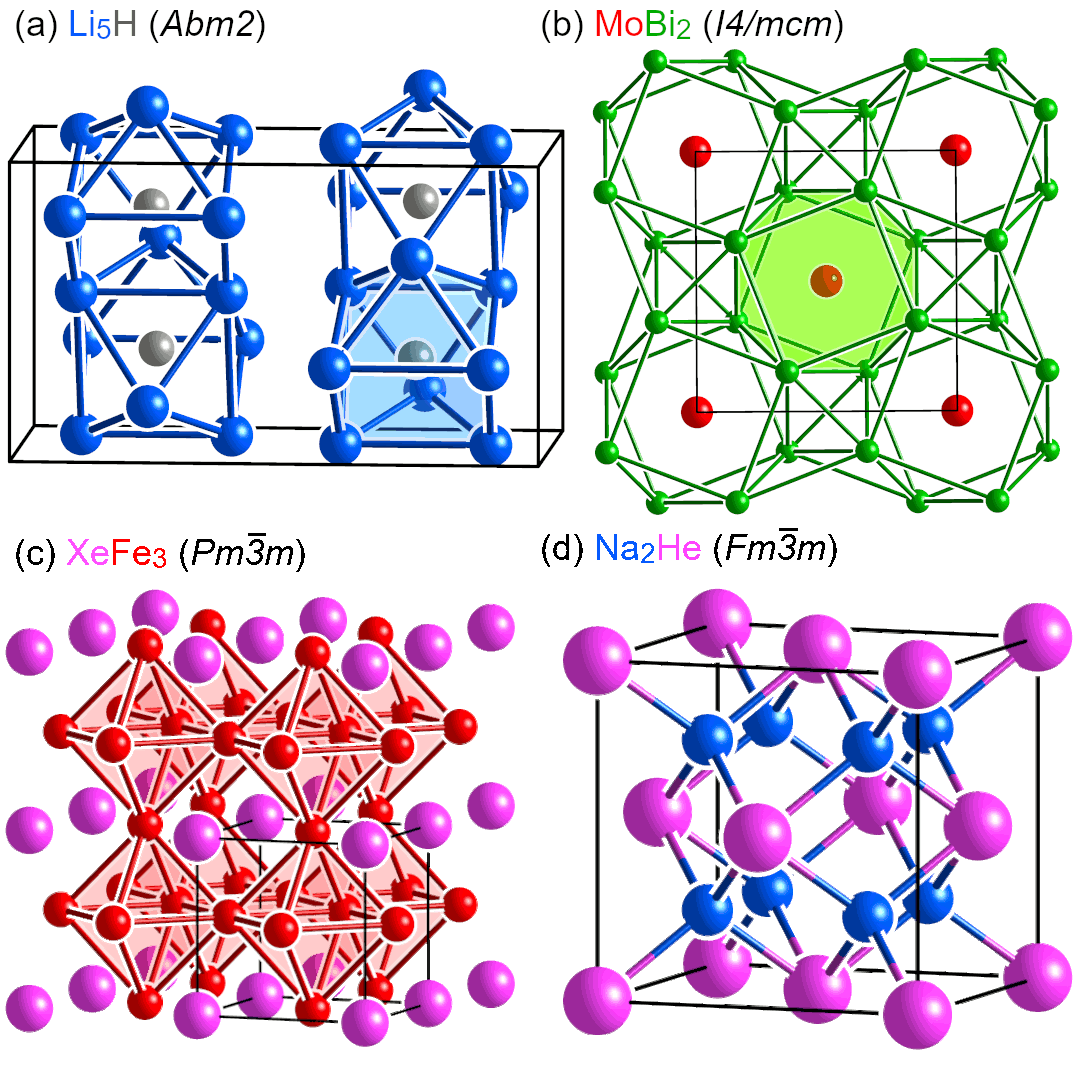}
\end{center}
\caption{Examples of strange stoichiometries that arise in compressed materials: (a) predicted Li$_5$H ($Abm2$) built from Li$_8$H subunits that behave as superalkali atoms, \cite{117} (b) synthesized MoBi$_2$ in the CuAl$_2$ type, with square antiprisms of Bi containing Mo atoms, \cite{Altman:2021} (c) synthesized $Pm\bar{3}m$ phase of XeFe$_3$, \cite{172,Stavrou:2018} and (d) synthesized $Fm\bar{3}m$ Na$_2$He, an insulating electride phase. \cite{Dong:2017}}
\label{fig:hpreview_fig3.png}
\end{figure}

A wide array of metal-hydrogen systems have been predicted to take on stoichiometries under pressure that at first may sound absurd. In the Li-H system, the LiH stoichiometry, which is the only one stable at 1~atm, is joined by LiH$_{2-8}$ \cite{109} and Li$_{4-9}$H \cite{117} phases under pressure. On the Li-rich side, two nearly isoenthalpic Li$_5$H structures with $Cmc2_1$ and $Abm2$ symmetry were predicted to be most enthalpically favorable. \cite{117} Both are built of Li$_8$H building blocks that can be viewed as distorted bicapped trigonal antiprisms with Li atoms on the vertices and H atoms at the center (Figure~\ref{fig:hpreview_fig3.png}a). The Li$_8$H block can be thought of as a superalkali atom with one more electron than a closed-shell number of eight, similar to building blocks in suboxide compounds such as Rb$_9$O$_2$ and Cs$_{11}$O$_3$. \cite{174} Since the electronegativity difference between Li and H is so large, electron transfer occurs from Li to H atoms, resulting in Li$^+$ cations and hydridic hydrogen, with the extra electrons donated by Li treated as anions to balance the charge. The interactions within the Li$_8$H cluster are treated as ionic, with metallic bonding on the outside.

On the the hydrogen-rich side of the Li-H system, several phases were found to be stable against decomposition into LiH and hydrogen, with LiH$_2$ and LiH$_6$ lying on the binary convex hull between 150 and 300~GPa. \cite{37,109} LiH$_2$ is semimetallic, with a lattice of Li$+$ cations and hydridic hydrogens combined with 1D chains of H$_2$ molecules. In LiH$_6$, electron transfer occurs from Li to H$_2$ units, such that each H$_2$ molecule carries $1/3$ of a negative charge. This partially fills the H$_2$ $\sigma^*$ antibonding bands, lengthening the intramolecular H-H distances. Motivated by theoretical predictions of the propensity of LiH$_6$ to become superconducting, \cite{175} experimentalists successfully synthesized a lithium polyhydride compound above 130~GPa in a DAC. \cite{176} It was suggested that Li diffuses into the the DAC during compression producing a layer of LiH$_6$ between the diamond and the sample, and another layer of LiH$_2$ between LiH$_6$ and LiH. The predicted metallicity was not detected in experiments up to 215~GPa -- however, later calculations using a van der Waals including density functional  suggested that LiH$_2$, LiH$_7$ and LiH$_9$ may be candidate structures for the synthesized phases. \cite{Chen:2017a}

Many other metal hydrides with unusual stoichiometries have been predicted within DFT for MH$_n$ phases, with M = alkali metal and $ n \geq 2$ or M = alkaline or rare earth metal and $ n > 2 $. \cite{37,98,110,111,112,113,114,115,116,177,178,Liu:2017-La-Y,100,Peng:2017a,ScH6-1,ScH6-2,ZrH6,Li:2019,Sun:2020}  A major motivation for these studies was the suggestion that ``chemical precompression", achieved by doping hydrogen with other elements, would lower hydrogen's metallization pressure and the subsequent onset of superconductivity. \cite{66,179} 

As with polyhydrides of Li, polyhydrides of Na were successfully synthesized in a DAC above 40~GPa and 2000~K. \cite{170} Experimental XRD patterns and Raman spectral peaks matched well with those predicted for NaH$_3$ and NaH$_7$ phases, which possessed H$_3^-$ units. \cite{110,170} Several alkaline metal hexahydrides (for example, MgH$_6$ at 300~GPa, \cite{MgH6} CaH$_6$ at 150~GPa, \cite{98} SrH$_6$ at 250~GPa, \cite{116,178} ScH$_6$ at 285, 300 and 350~GPa, \cite{ScH6-1,ScH6-2,Peng:2017a} YH$_6$ at 120~GPa, \cite{YH6-1,Peng:2017a,100} ZrH$_6$ at 295~GPa, \cite{ZrH6} and LaH$_6$ at 100~GPa \cite{Peng:2017a}) have been predicted to be good metals with almost free-electron like DOS at the Fermi level. Therefore, it is not surprising that they were predicted to be superconducting with relatively high estimated $T_c$ values. Even higher $T_c$s have been predicted (and some have been observed) for the decahydrides YH$_{10}$ \cite{Liu:2017-La-Y,Peng:2017a} and LaH$_{10}$,\cite{Liu:2017-La-Y,Peng:2017a,geballe2018synthesis,Pers105,drozdov2019superconductivity} as well as in CeH$_9$ \cite{Li:2019} and YH$_9$.\cite{Kong:2021} 

Several elemental systems that are immiscible at 1~atm can form alloys under pressure. Li-Be is one of these, with no known binary alloys at ambient pressure. However, theoretical studies have predicted that between 20-200~GPa, several binary Li-Be phases including Li$_3$Be, LiBe, LiBe$_2$, and LiBe$_4$ become thermodynamically stable. \cite{68} Their stabilization is proposed to occur through a Hume-Rothery mechanism of Fermi sphere-Brillouin zone intersection.\cite{Degtyareva:2006} A detailed analysis on the $P2_1/m$ LiBe phase reveals a unique structure comprised of alternating layers of puckered Li square nets and Be triangular nets. The two-dimensionality of the Be layers -- unexpected in a bimetallic alloy -- is seen in a sharp, step-like peak in the electronic DOS at 82~GPa. The 2$s$ electrons of Li join the Be layers as the Li 1$s$ cores start to overlap, but the Be 1$s$ core, due to their smaller size do not. It was proposed that the LiBe alloy, therefore, may be viewed as a manifestation of a Zintl-type compound. \cite{166} 

Bismuth is a component of several widely studied topological materials \cite{Bi2Te3,Na3Bi} and superconductors \cite{Reynolds:1950,Matthias:1966}, but is immiscible at ambient pressure with many transition metals -- to the point that bismuth has been used as a flux in ambient-pressure syntheses. \cite{Thompson:2011} Under pressure, the situation changes substantially and binary phases of Bi with Fe \cite{Walsh:2016,Amsler:2017}, Co \cite{Schwarz:2013}, Cu \cite{Clarke:2016, Clarke:2017,Amsler:2017b}, Mo \cite{Altman:2021}, and Sn \cite{Amsler:2018} have all been predicted and in some cases synthesized. MoBi$_2$,\cite{Altman:2021}  FeBi$_2$, \cite{Walsh:2016,Amsler:2017} MnBi$_2$, \cite{Walsh:2019} and NiBi$_2$ \cite{Amsler:2018} (the Mn-Bi and Ni-Bi systems display some ambient-pressure miscibility) all crystallize in the CuAl$_2$ structure type shown in Figure~\ref{fig:hpreview_fig3.png}b that is adopted by several transition metal pnictides. Several of these phases are predicted \cite{Amsler:2017,Amsler:2017b} or measured \cite{Clarke:2016,Schwarz:2013} to be superconducting.

A number of phases containing noble gases, which are highly unreactive at 1~atm, can be stabilized under pressure. This may be key to understanding why the abundance of xenon on earth is much lower than would be expected based on its abundance in interstellar space, otherwise known as the ``missing xenon paradox”.  \cite{180} Many studies have been carried out to solve this paradox; \cite{181} some hypothesizing that the``missing xenon” could react with other elements under pressure. 

Theoretical studies have predicted that Fe, Ni and O can form thermodynamically stable structures with Xe under pressure. \cite{122,172,173} Both Xe-Fe and Xe-Ni phases were predicted to be stable above 200 and 250~GPa, respectively, with their high-temperature stabilities validated under the quasiharmonic approximation. \cite{172} Later experimental studies detected both XeFe$_3$ (Figure~\ref{fig:hpreview_fig2.png}c) and XeNi$_3$ alloys, as well as a mixed iron-nickel xenon compound. \cite{Dewaele:2016b,Stavrou:2018}

Additionally, a plethora of Xe-O phases including Xe$_2$O, Xe$_{1, 3, 7}$O$_2$, and XeO$_3$ were computed to be stable against decomposition into the component elements above 75~GPa. \cite{122,173} These studies suggest that the “missing Xe” might be trapped by the minerals containing oxygen that exist in the Earth's lower mantle. \cite{173} A recent theoretical study found that Xe can react with the newly discovered iron peroxide compound (FeO$_2$) to form ternary compounds of xenon, iron, and oxygen with stoichiometries of Xe$_2$FeO$_2$ and XeFe$_3$O$_6$ predicted at lower mantle conditions. \cite{Xe-Fe-O} Xe-O interactions were found to be strong in these ternary compounds. No other noble gas elements were reactive with iron peroxide, suggesting that it might specifically be “blamed” for the missing xenon. Several other elements have been predicted to form compounds with xenon under pressure, including magnesium, \cite{182}  nitrogen, \cite{Xe-N} cesium \cite{Zhang:2015}, and yttrium, \cite{Zhou:2019} while compound formation has been experimentally observed when xenon is combined with water \cite{183} and hydrogen \cite{184}. 

Other noble gases are known to become reactive under pressure -- compounds have been predicted for Ar with Ni \cite{Adeleke:2019}, Li \cite{Li:2015}, and Mg \cite{182}. Particularly exciting is a synthesized Na$_2$He phase stable above 113~GPa in the fluorite structure (Figure~\ref{fig:hpreview_fig2.png}d) \cite{Dong:2017}. Subsequent theoretical studies uncovered compounds of Fe and He \cite{Monserrat:2018}, He and ammonia \cite{Liu:2020}, and a ternary FeO$_2$He compound \cite{Zhang:2018}. One study investigated the stability of alkali metal oxides or sulfides intercalated with He and Ne. \cite{Gao:2019} It has been proposed that helium can serve as something of an inert ``spacer" in ionic compounds with unequal numbers of cations and anions. \cite{Liu:2018} Repulsions between the majority ions can prevent the formation of close-packed structures -- in which case an added He atom provides relief in the $PV$ energetic term -- and the added He keeps the majority ions separated from one another. Through this lens, the Na$_2$He phase, an electride, can be treated as a compound of (Na$^+$)$_2$$\cdot2e^-$ with He.

\subsection{Structure and Bonding}\label{subsec11}

Increased densification promotes electron density overlap between neighboring atoms, which may react by adopting electron rich or electron poor multi-centered bonding schemes. \cite{Grochala:2007a} In his seminal work on chemical bonding, Pauling predicted that sufficiently large pressure would shorten the intermolecular hydrogen bonds in water molecules and elongate the intramolecular H-O bonds. \cite{129} At some point, those two types of H-O bonds would then become equivalent, using the lone pairs on an oxygen atom to form bonds with the neighboring hydrogen atoms that are hydrogen-bound to oxygen at ambient pressure. This phenomenon has been observed in experiments for the ice X phase above 60~GPa, whose intra and intermolecular O-H bonds are symmetrized and form three dimensional networks resembling the diamond structure, leaving oxygen atoms four-coordinated. \cite{158,159}  Several other hydrogen bond-containing systems have been found both experimentally and theoretically to show similar bond symmetrizing behavior under pressure, including HF, HCl, HBr, \cite{160} and the metastable phase of Li-F-H. \cite{LiFHn} 


Interesting structural motifs where multicentered bonds are formed are present in several predicted high pressure hydride phases. \cite{103,111,112,115,Duan:2015,LiFHn} For example, the trihydrogen cation (H$_3^+$), a dominant interstellar molecule, maximizes the overlap of hydrogen bonding orbitals to form three-center two-electron bonds taking the shape of an equilateral triangle. The symmetric H$_3^+$ cluster occurs in the H$_5$Cl phase that has been theoretically predicted to be stable from 60-300~GPa. \cite{103,Duan:2015}. Another ternary hydride phase that is metastable at 300~GPa, LiF$_4$H$_4$, contains a distorted version of this motif. \cite{LiFHn} 

With two more electrons than the trihydrogen cation, the trihydrogen anion (H$_3^-$) has a linear geometry with three-center four-electron bonds. The ground state configuration of the trihydrogen anion, based on \textit{ab initio} calculations, shows a large difference between the two H-H bonds: 2.84~\AA{} versus 0.75~\AA{}. \cite{163} The PES of this cluster has two wells containing two equivalent configurations, with the transition state at the peak of the energy barrier corresponding to the bond-symmetrized arrangement where both H-H distances are 1.06~\AA{}. Hydrides of heavy group I (i.e. K, \cite{112} Rb, \cite{111} and Cs \cite{113}) or group II (i.e. Ba \cite{115}) metals display symmetric H$_3^-$ ions. For example, several nearly isoenthalpic CsH$_3$ structures, all of which are built from Cs$^+$ and H$_3^-$, become the lowest enthalpy points on the convex hulls above 30~GPa. \cite{113} Computational experiments have shown that the softness of the cation is related to the formation of symmetric H$_3^-$ anions under pressure. \cite{112}

\begin{figure}
\begin{center}
\includegraphics[width=1\columnwidth]{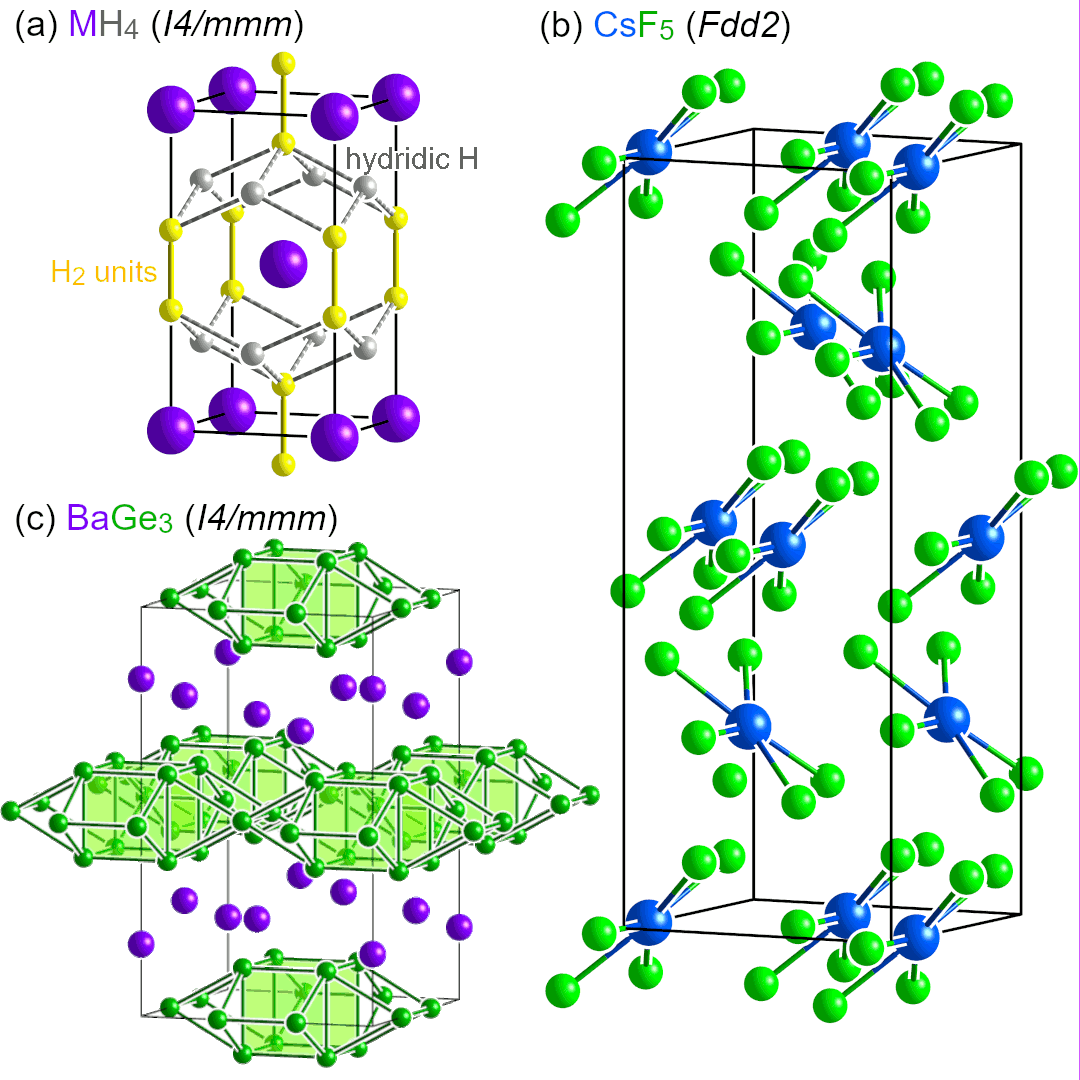}
\end{center}
\caption{Structural motifs in materials at high pressures: (a) predicted and synthesized MH$_4$ $I4/mmm$ phases adopted by several alkaline earth and rare earth hydrides comprised of hydridic H (grey) and quasimolecular H$_2$ (yellow) units; \cite{MH4} (b) predicted CsF$_5$ ($Fdd2$) phase with lightly distorted planar pentagonal coordination environment of Cs reminiscent of the XeF$_5$ molecule; \cite{102} (c) predicted and synthesized BaGe$_3$ ($I4/mmm$) phase adopted by several polar intermetallic compounds with polyhedra built up of tetrel dumbbells. \cite{168,169}}
\label{fig:hpreview_fig4.png}
\end{figure}

The high pressure structures of metal hydrides present numerous curious bonding arrangements for hydrogen atoms. In the pressure-stabilized alkaline earth or rare metal tetrahydride MH$_4$ phases, \cite{98,116,178,Qian:Sc-2017,ScH6-1,ScH6-2,Peng:2017a,Liu:2017-La-Y,Li:2019,Zurek:2018c} pressure can also induce elongation or even dissociation of bonds. \cite{MH4} These tetrahydrides assume the same $I4/mmm$ structure (ThCr$_2$Si$_2$-type) that occurs in over 700 AB$_2$X$_2$ compounds and consists of three parts: metal cations, hydridic hydrogen, and quasi-molecular hydrogen units (Figure~\ref{fig:hpreview_fig4.png}a). Charge transfer from trivalent or tetravalent metal atoms to the H$_2$ $\sigma _{u^*}$ bands can elongate or break the H-H bonds in the quasi-molecular units. Any still existing H-H bonds are further weakened by a Kubas-like interaction of electron transfer from the H$_2$ $\sigma_{g}$ to the metal $d$ orbitals, and metal $d$ orbitals to the H$_2$ $\sigma_{u^*}$  states. This means that the H-H distance -- which is related to the phases' propensity towards superconductivity -- within the quasi-molecular orbitals can be tuned by the choice of metal atoms and the applied pressure. \cite{MH4}

Other wild and wonderful hydrogenic motifs are plentiful among the metal hydrides; among the most iconic are the ``clathrate-like" cages appearing in superconducting phases such as CaH$_6$. \cite{98}  Isostructural to YH$_6$, \cite{Liu:2017-La-Y,Peng:2017a,100} MgH$_6$, \cite{MgH6} ScH$_6$, \cite{ScH6-1,ScH6-2} and more, the body-centered lattice of CaH$_6$ features Ca atoms at the centers of an H$_{24}$ sodalite-like framework of six square and eight hexagonal faces. Electron donation from the Ca atom to the hydrogen sublattice has been postulated to populate a half-filled degenerate nonbonding state for an H$_4$ square unit, precluding condensation into H$_2$ molecular units but maintaining weak bonding character. \cite{98} Similar weakly bound clathrate cages appear in MH$_9$ phases such as CeH$_9$ \cite{Li:2019} and MH$_{10}$ compounds including LaH$_{10}$. \cite{Liu:2017-La-Y} The Sc-H family of high-pressure phases displays another strange structural motif: H$_5$ pentagons. \cite{ScH6-1,Xie:2020} These curious units appear joined together in strips in the ScH$_9$ and ScH$_{12}$ structures, where plots of the Electron Localization Function (ELF) confirms the presence of localized bonding. \cite{ScH6-1} In ScH$_{10}$ at $\sim$200~GPa two nearly isoenthalpic phases with $Immm$ and $P6_3/mmc$ symmetry feature H$_5$ pentagons fused to one another in sets of three in a ``pentagraphene-like" motif. \cite{Xie:2020} As with many other metal superhydride phases, the electrons donated from the metal atom (investigated for a hafnium analogue HfH$_{10}$) go into antibonding orbitals of the hydrogen framework, weakening the H-H bonding and reducing H$_2$ molecular character.

Pressure-induced changes in the energy ordering of atomic orbitals can cause orbitals that would not normally participate in bonding to do so. One good example is CsH$_3$, where the Cs adopts a Cs$^+$ configuration and the energy of the Cs 5$p$ orbitals increases under pressure so they interact with the hydrogen $s$ electrons. \cite{113}. In the CsF$_{2-6}$ phases found above 10~GPa that are stable against decomposition into CsF and F$_2$, the geometries and structural motifs mirror the isoelectronic XeF$_n$ molecules. \cite{102} For example, the building blocks of CsF$_5$ (Figure~\ref{fig:hpreview_fig4.png}b) are pentagonal planar with Cs as the center atoms coordinated by five F atoms, as seen in XeF$_5$ molecules. \cite{102} At 100~GPa and high fluorine content (F:Cs ratio $\geq$ 2), the oxidation state of Cs can increase almost linearly past $+1$ with increasing F content, according to Bader analyses. In CsF$_2$ and CsF$_3$, covalent bonding was indicated between the Cs semicore 5$p$ and F 2$p$ orbitals based on a crystal orbital Hamiltonian population (COHP) analysis. \cite{165} Thus, the pressure-reordering of orbital energies has serious consequences on the possible bonding interactions between cesium and fluorine atoms, allowing inner-shell electrons to actively participate.

A series of polar intermetallic compounds MA$_3$, where M is an alkaline earth or rare earth metal and A is a group 14 tetrel element, adopt an $I4/mmm$ structure that only becomes stable under high pressure. In this structure, found in CaGe$_3$, \cite{162} CaS$_3$, YSi$_3$, and LuSi$_3$ \cite{167}, and predicted prior to its synthesis for BaGe$_3$ \cite{168,169}, the tetrel atoms form a condensed lattice of  A-A dumbbells, which is layered with the metal atoms (Figure~\ref{fig:hpreview_fig4.png}c). Investigations using the electron localizability indictor or the COHP show covalent bonding within the tetrel dumbbells and multicenter interactions between the metal atom and the tetrel sublattice. \cite{162,167,169} Several of the phases were predicted to become Bardeen-Cooper-Schrieffer (BCS) conventional superconductors below 10~K according to first-principles calculations. \cite{162,167,169} Furthermore, some of the phases could be recovered at ambient conditions. The strong bonding networks formed under pressure protect against decomposition as the pressure is released, since large amounts of energy would be required to break the bonds. In this way, high-pressure high-temperature techniques can be successfully used to synthesize compounds with increased coordination components that do not align with electron-precise counting schemes such as the Zintl concept. \cite{166}.

\subsection{Superconductivity}\label{subsec12}

The propensity for superconductivity is one of the properties that is significantly affected by the pressure variable. At least 23 elements undergo superconducting transitions under pressure, \cite{185} and the $T_c$ itself depends greatly on pressure for superconducting elements and compounds alike. The material with the highest 1~atm $T_c$, HgBa$_2$Ca$_2$Cu$_3$O$_8$, is an unconventional superconductor whose $T_c$ of 133~K can be increased further to 164~K at 31~GPa. \cite{186} For over 20 years, this compound had the highest critical temperature known, until a record-breaking 203~K was measured at 150~GPa in the sulfur-hydrogen system. \cite{189} 

Hydrogen-rich compounds have become a target for pursuing high temperature superconductivity under pressure as a proxy for metallic hydrogen itself. The reason for this stems from Ashcroft's proposal that the addition of another element to hydrogen could reduce the required pressure for transitioning into the superconducting state through ``chemical pre-compression". \cite{136,66,DopingH} Herein, we overview several recent examples of high pressure hydrides with remarkable $T_c$ values, and also illustrate how the successful synergy between theoretical predictions and experimental synthesis led to their discovery.

Previous theoretical studies \cite{187} that estimated the superconducting properties of compounds with the H$_2$S stoichiometry at high pressure using the Allen-Dynes-modified McMillan equation \cite{188} identified a phase with an estimated $T_c$ of 80~K at 160~GPa. Motivated by this prediction, experimental work proceeded, resulting in a phase with measured $T_c$ below 100~K in good agreement with the expected values. \cite{189} However, when the sample was prepared at higher temperatures above 300~K, a material with a record-breaking $T_c$ value of 203~K at 150~GPa was synthesized instead. \cite{189}  

\begin{figure}
\begin{center}
\includegraphics[width=1\columnwidth]{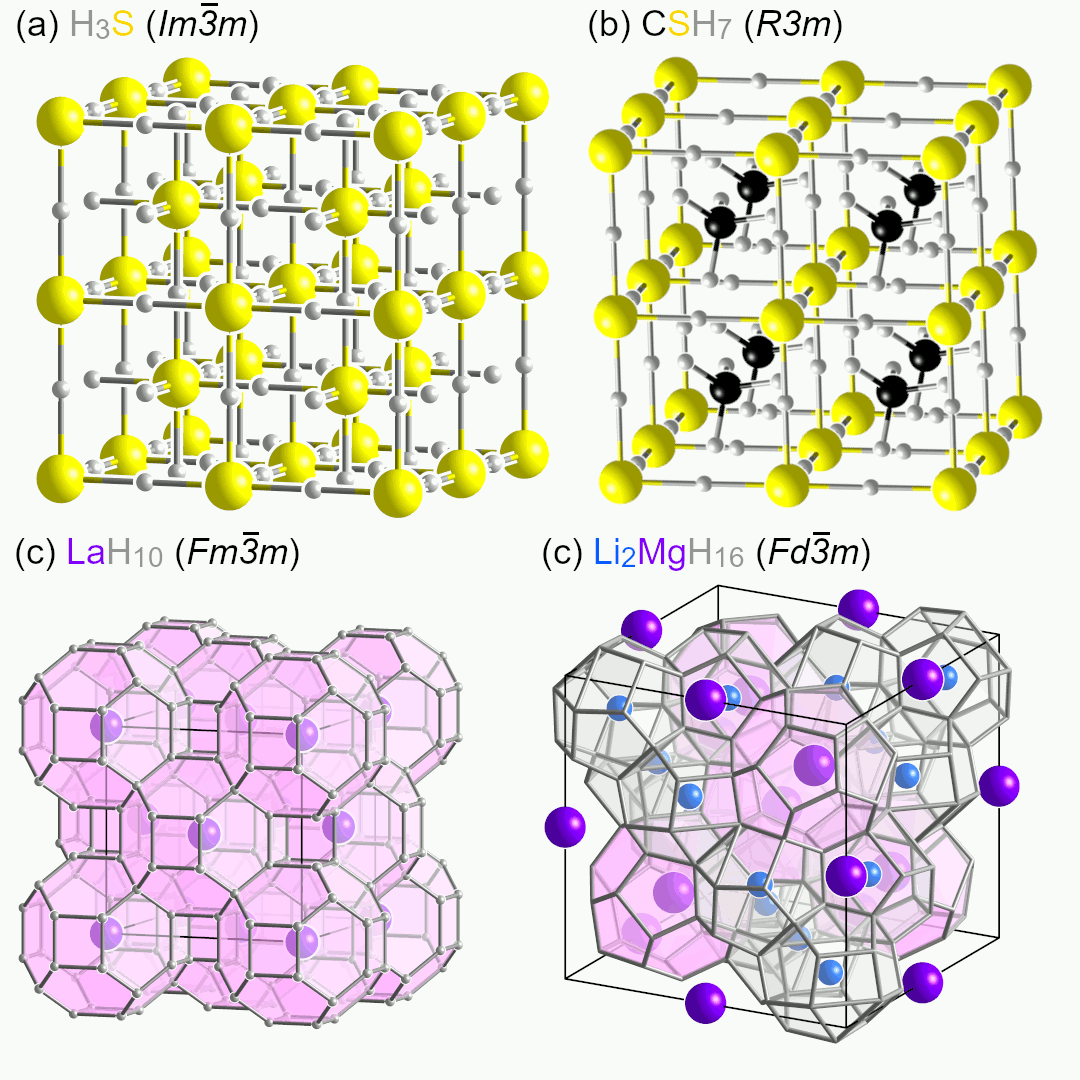}
\end{center}
\caption{Selected high-temperature superconducting hydrides that have been calculated to be stable or metastable at high pressures: (a) $Im\bar{3}m$ H$_3$S, the first high-temperature superconducting hydride that was synthesized,~\cite{189,Duan:2014} (b) $R3m$ CSH$_7$, one example of a predicted ternary superconducting hydride,~\cite{CSH7} (c) LaH$_{10}$, the first clathrate-like superhydride to be made,~\cite{drozdov2019superconductivity,Pers105} and (d) Li$_2$MgH$_{16}$, a predicted ternary superhydride with an immensely high predicted $T_c$ of 473~K~\cite{Li2MgH16}}
\label{fig:hpreview_fig5.png}
\end{figure}

In a parallel set of studies, another group of researchers had previously synthesized an (H$_2$S)$_2$H$_2$ phase above 3.5~GPa. \cite{Strobel:2011a} This work inspired a follow-up theoretical investigation that identified a  series of structures with the H$_3$S stoichiometry at higher pressures, including an $R3m$ phase with a $T_c$ of 155-166~K at 130~GPa.  \cite{Duan:2014} At even higher pressures, this structure was predicted to transform to a related $Im\overline{3}m$ phase (Figure~\ref{fig:hpreview_fig5.png}a) whose $T_c$ was estimated as 191-204~K at 200~GPa -- in line with the experimental results of Ref.\ \cite{189}. Subsequent XRD analysis suggested that the superconducting sample contained a mixture of the two predicted phases along with the $\beta$-polonium phase of sulfur. \cite{192} The high $T_c$ was later confirmed by the Meissner effect, \cite{MeissnerH3S} and an optical reflectivity experiment categorized the hydrogen sulfide material as a conventional BCS superconductor whose superconductivity arises from electron-phonon coupling. \cite{OptRefH3S} 

The exciting work on H$_3$S prompted a slew of investigations into the compositions and structures of the superconducting phases, the factors that contribute to $T_c$, the isotope effect, the importance of anharmonicity, and the role of hydrogen's quantum nature. \cite{a324,a357,91,194,195,196,197,198,a373,CSH38,a375,a376,a377,a378,a379,a380,a381,a382,a383,Bianco:2018,Amsler:2019} For example, one theoretical investigation concluded that the decomposition of the $Im\overline{3}m$ phase into (SH$^-$)(H$_3$S$^+$) perovskites is favorable. \cite{a384} First-principles molecular dynamics (MD) simulations at 200~GPa and 200~K suggested that this perovskite phase can be further separated into tetragonal and cubic regions, and the resulting MD structure produces an XRD pattern in good agreement with the experimental observations. \cite{a385} 

The high $T_c$ of H$_3$S phases also inspired multiple studies on ternary systems based on adding a third element to the S-H system. The electronic structure of H$_3$S features a maximum in the electronic DOS near the Fermi level due to the presence of a pair of van Hove singularities on either side. \cite{CSH38,a378} Tuning the electronic structure via doping with another element could shift the Fermi level to lie directly on the maximum, strengthening the electron-phonon coupling and raising the $T_c$ of the system. 

In one study, currently in the spotlight, a carbonaceous sulfur hydride with a measured $T_c$ of 288~K at 267~GPa was reported. \cite{CSH9} More recent XRD analysis suggests that this phase is a derivative of the Al$_2$Cu type. \cite{CSH10,CSH11} Before the experimental discovery of the carbonaceous sulfur hydride, structure predictions had actually been carried out on various carbon-doped sulfur hydride stoichiometries: C$_{1-2}$S$_{1-2}$H$_{1-8}$ up to 300~GPa, \cite{CSH7} and C$_{1-4}$S$_{1-4}$H$_{1-36}$ at 100~GPa. \cite{CSH35} A stoichiometry of CSH$_7$ emerged as particularly interesting in the calculations. A number of metastable phases arising from the intercalation of methane molecules into the H$_3$S framework (Figure~\ref{fig:hpreview_fig5.png}b) were found. The estimated $T_c$s for the predicted CSH$_7$ structures fell well below that measured for the carbonaceous sulfur hydride (194~K at 150~GPa and 181~K at 100~GPa), and their calculated equations of state also did not match the experimental C-S-H data. A first-principles investigation using the virtual crystal approximation (VCA), which employs pseudoatoms based on weighted potentials of different elements to model doping, found that a small amount of carbon incorporation into the H$_3$S lattice was sufficient to increase the estimated $T_c$ to 288~K. \cite{CSH36} However, the VCA method disregards local chemical structure and bonding. Therefore, a more recent theoretical study was carried out on supercells of H$_3$S doped with 1.85-25\% carbon (interpreted as SH$_3 \rightarrow$ CH$_3$, producing six- or four-coordinate C, or SH$_3 \rightarrow$ CH$_4$, producing four-coordinate C). \cite{CSHarXiv} Contrary to results obtained with the VCA, it was found that the doping lowers the DOS at the Fermi level, but increases the logarithmic average phonon frequency. The most promising phase for which $T_c$ could be explicitly computed, CS$_3$H$_{13}$, was predicted to possess essentially the same $T_c$ as $R3m$ H$_3$S at 270~GPa. It was postulated that the most promising stoichiometries for high-$T_c$ were CS$_{15}$H$_{49}$ or CS$_{53}$H$_{163}$, but explicit calculations could not be performed due to the system size.

Besides sulfur, hydrides of another $p$-block element have inspired interest. After loading phosphine into a DAC, researchers observed the onset of superconductivity at 30 and 103~K at 83 and 207~GPa, respectively. \cite{193} Follow-up theoretical studies appeared quickly, since the structural and electronic properties of the superconducting phosphorus compounds were not fully characterized. \cite{90,118,201,202,Bi-PH2}. Predictions suggested that compression caused phosphine to decompose, with the measured superconductivity arising from some combination of PH, PH$_2$, PH$_3$, and other metastable phosphorus hydrides. \cite{90,118,202,Bi-PH2} In one example, an $I4/mmm$ PH$_2$ phase had an estimated $T_c$ of 70~K at 200~GPa, \cite{118}, increasing to 78~K at 220~GPa. \cite{90} At lower pressures, PH$_2$ phases were found to be metastable, with enthalpies of formation and $T_c$ values in line with the experimental observation at 83~GPa. \cite{Bi-PH2} A unique structure comprised of cubic-like phosphorus layers capped by hydrogen atoms at the top and bottom layers and further separated by layers of molecular hydrogen was observed. Interestingly, the superconductivity was mainly derived from the P-H layers, while the neutral and mobile H$_2$ layers served to stabilize the structure by separating the negatively charged H atoms. 

Meanwhile, prior to the discovery of the carbonaceous sulfur hydride, the high $T_c$ record of H$_3$S was broken by another class of high pressure hydrides: binary combinations of hydrogen with alkaline and rare earth metals with the hydrogen atoms arranged in clathrate-like cages, as described in the previous section. \cite{drozdov2019superconductivity,Pers105} Experimental measurements on $Fm\bar{3}m$ LaH$_{10}$ detected $T_c$ values as high as 250~K \cite{drozdov2019superconductivity} and 260-280~K \cite{Pers105}. 

Theoretical investigations had previously predicted the stability of this phase (Figure~\ref{fig:hpreview_fig5.png}c) with La atoms occupying fcc positions on a cubic lattice surrounded by an H$_{32}$ clathrate-like cage. ~\cite{Liu:2017-La-Y,Peng:2017a} An isotypic yttrium analogue has also been predicted, with a $T_c$ of 305-325~K at 250~GPa calculated via numerical solution of the Eliashberg equations, \cite{Liu:2017-La-Y} but such a phase has not yet been experimentally obtained.  A ternary structure synthesized by mixing La and Y into the $Fm\bar{3}m$ lattice has demonstrated $T_c$ up to 253~K at 183~GPa, where the addition of La is thought to stabilize the YH$_{10}$ framework. \cite{Semenok:2021b} In addition, other ternary hydrides with impressive predicted $T_c$s and pressure stability ranges have been identified, including a metastable Li$_2$MgH$_{16}$ phase (Figure~\ref{fig:hpreview_fig5.png}d) that is predicted to be a ``hot" superconductor with a $T_c$ $\sim$473~K at 250~GPa. \cite{Li2MgH16} A family of phases based on adding a third element into void spaces of the MH$_{10}$ lattice, resulting in stoichiometries such as LaBH$_8$ \cite{DiCataldo:2021,Liang:2021} and KB$_2$H$_8$ \cite{Gao:2021} show promise in maintaining stability to pressures below a megabar. These discoveries are paving the way to warm and light superconductivity at more readily accessible pressures.

\section{Conclusion}

For the vast majority of us who have never left Earth's surface, the only way in which we adjust our lives to account for the affects of pressure is to adapt cooking recipes to account for water boiling at a lower temperature at high altitudes. From the top of Mount Everest to the bottom of the Mariana Trench the pressure difference is roughly only a tenth of a gigapascal. The development of experimental techniques to access successively higher pressures, coupled with a dramatic improvement in theoretical methods and computational power has allowed researchers to explore materials at extreme high pressure conditions. In many cases this has led to the discovery of a profoundly different chemistry.

High pressure alters the relative orbital energies of atoms, driving
electronic transitions within atoms and ultimately resulting in unfamiliar, let us call it extreme, chemistry. In some cases, valence electrons compressed into void regions become interstitial quasiatoms, while in
others the energies of core orbitals increase enough to partake in bonding
and chemistry. Elements that are stubbornly immiscible with one another, or
are just plain unreactive, under atmospheric conditions form compounds at
high pressures. Unfamiliar structures can emerge, from extended 3D networks
of hydrogen~\cite{98} to lattices of tetrel dumbbells~\cite{162} to
pentagonal planar CsF$_5$ units~\cite{113} (with Cs taking on a charge $>
+1$ to boot!).

Experimental methods to synthesize and probe such strange phases have
advanced greatly, with improvements in diamond anvil cells including
toroidal~\cite{Jenei:2018,Dewaele:2018} and double-stage diamond anvil
cells~\cite{Lobanov:2015}, as well as in dynamic compression
experiments~\cite{Fratanduono:2021,Hansen:2021} accessing higher and higher
pressures. Theoretical methods remain closely intertwined with experiment, providing
guidance on promising systems to investigate and elucidating the structural
and electronic character of the phases that are found. 

Crystal structure prediction (CSP) techniques have proven immensely useful in helping researchers begin to develop chemical intuition under pressure. Ranging from methods that finely 
explore particular regions of a PES (e.g.\ metadynamics and minima hopping) to those that promote broad exploration (e.g.\ random searches, particle swarm optimization, and genetic algorithms), CSP has been used to identify a plethora of new phases. The synergy between experiment and theory is evident in the rapid progress made in the field of high-temperature superconductivity in the hydrides, beginning with the discovery of the record-breaking H$_3$S and LaH$_{10}$ phases. 
Experimental reports of room-temperature superconductivitiy in a C-S-H phase has inspired numerous studies. Complex phases combining three, four, or even more elements and strange chemical behavior await to be uncovered as researchers further explore this fascinating field.

\backmatter

\bmhead{Acknowledgments}

We acknowledge the NSF (DMR-1827815) for financial support.  This material is based upon work supported by the U.S.\ Department of Energy, Office of Science, Fusion Energy Sciences under Award Number DE-SC0020340 to E.Z. K.P.H.\ thanks the US Department of Energy, National Nuclear Security Administration, through the Chicago-DOE Alliance Center under Cooperative Agreement Grant No.\ DE-NA0003975 for financial support.

\bibliography{review-bibliography}


\end{document}